\documentclass[aps,prl,reprint,superscriptaddress,showpacs]{revtex4-1}
\usepackage{graphicx}% Include figure files
\usepackage{amsmath}
\usepackage{color}
\begin{document}

%\preprint{}

\title{Superconducting tunneling spectroscopy of spin-orbit coupling\\ and orbital depairing in Nb:SrTiO$_3$}


\author{Adrian G. Swartz}
\email[]{aswartz@stanford.edu}
%\thanks{These authors contributed equally.}
\affiliation{Geballe Laboratory for Advanced Materials, Stanford University, Stanford, California 94305, USA}
\affiliation{Stanford Institute for Materials and Energy Sciences, SLAC National Accelerator Laboratory, Menlo Park, California 94025, USA}
\affiliation{Department of Applied Physics, Stanford University, Stanford, California 94305, USA}

\author{Alfred K. C. Cheung}
%\email[]{}
%\thanks{}
\affiliation{Department of Physics, Stanford University, Stanford, California 94305, USA}

\author{Hyeok Yoon}
%\thanks{These authors contributed equally.}
\affiliation{Geballe Laboratory for Advanced Materials, Stanford University, Stanford, California 94305, USA}
\affiliation{Stanford Institute for Materials and Energy Sciences, SLAC National Accelerator Laboratory, Menlo Park, California 94025, USA}
\affiliation{Department of Applied Physics, Stanford University, Stanford, California 94305, USA}

\author{Zhuoyu Chen}
%\thanks{These authors contributed equally.}
\affiliation{Geballe Laboratory for Advanced Materials, Stanford University, Stanford, California 94305, USA}
\affiliation{Stanford Institute for Materials and Energy Sciences, SLAC National Accelerator Laboratory, Menlo Park, California 94025, USA}
\affiliation{Department of Applied Physics, Stanford University, Stanford, California 94305, USA}

\author{Yasuyuki Hikita}
%\thanks{}
\affiliation{Stanford Institute for Materials and Energy Sciences, SLAC National Accelerator Laboratory, Menlo Park, California 94025, USA}

\author{Srinivas Raghu}
%\email[]{}
%\thanks{}
\affiliation{Stanford Institute for Materials and Energy Sciences, SLAC National Accelerator Laboratory, Menlo Park, California 94025, USA}
\affiliation{Department of Physics, Stanford University, Stanford, California 94305, USA}

\author{Harold Y. Hwang}
%\email[]{}
%\thanks{}
%
\affiliation{Geballe Laboratory for Advanced Materials, Stanford University, Stanford, California 94305, USA}
\affiliation{Stanford Institute for Materials and Energy Sciences, SLAC National Accelerator Laboratory, Menlo Park, California 94025, USA}
\affiliation{Department of Applied Physics, Stanford University, Stanford, California 94305, USA}


\date{\today}

\begin{abstract}
We have examined the intrinsic spin-orbit coupling (SOC) and orbital depairing in thin films of Nb-doped SrTiO$_3$ by superconducting tunneling spectroscopy. The orbital depairing is geometrically suppressed in the two-dimensional limit, enabling a quantitative evaluation of the Fermi level spin-orbit scattering using Maki's theory. The response of the superconducting gap under in-plane magnetic fields demonstrates short spin-orbit scattering times $\tau_{so} \leq 1.1$ ps. Analysis of the orbital depairing indicates that the heavy electron band contributes significantly to pairing. These results suggest that the intrinsic spin-orbit scattering time in SrTiO$_3$ is comparable to those associated with Rashba effects in SrTiO$_3$ interfacial conducting layers and can be considered significant in all forms of superconductivity in SrTiO$_3$.  
\end{abstract}

%\keywords{}

\maketitle


The relativistic spin-orbit interaction is fundamental in the solid state, connecting the conduction electron spin to the atomic, electronic, orbital, and structural symmetry properties of the material \cite{Winkler:2003}. 
SrTiO$_3$ is an oxide semiconductor with highly mobile $t_{2g}$ conduction electrons and exhibits superconductivity at the lowest known carrier density of any material \cite{Mazin:2011, Kim:2011, Lin:2013}. 
The relevance of the intrinsic spin-orbit coupling (SOC) for superconductivity in the bulk material remains an open question: the atomic SOC produces a relatively small splitting (29 meV \cite{Mazin:2011}) of the $t_{2g}$ bands, butmight be an important energy scale considering the small superconducting gap in SrTiO$_3$.  
Moreover, SrTiO$_3$ is the host material for unconventional two-dimensional (2D) superconductors such as FeSe/SrTiO$_3$ \cite{Qing:2012}, $\delta$-doped SrTiO$_3$ \cite{Kim:2011}, and LaAlO$_3$/SrTiO$_3$ \cite{Caviglia:2008}. 
Spin-orbit coupling in SrTiO$_3$ interfacial accumulation layers has been extensively studied both experimentally and theoretically \cite{Ben:2010, Caviglia:2010b, Nakamura:2012, Zhong:2013, Khalsa:2013, King:2014, Rout:2017}.
In these systems, Rashba SOC has been suggested to give rise to many of the unusual normal- and superconducting-state properties due to the broken inversion symmetry and the highly asymmetric confinement potential. 
Understanding the competition between the intrinsic and Rashba coupling scales is critical to understanding the spin-orbit textures and superconducting phases in both bulk and 2D systems.

%%%%%%%%%%%%Figure 1
%%%%%%%%%%%%
\begin{figure}[]
\includegraphics[width=85mm]{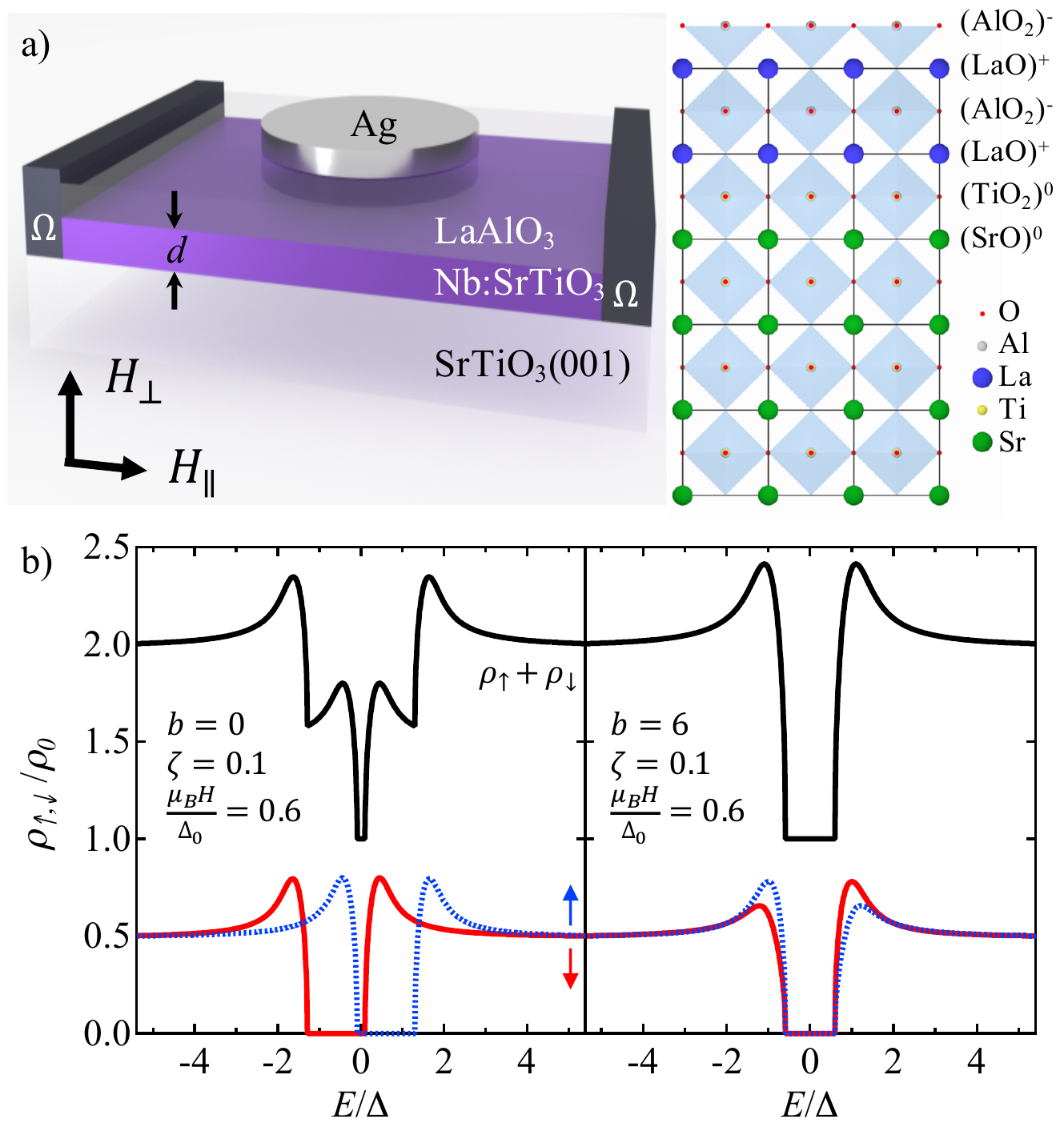}
\caption{\label{fig1} 
(Color online)
a) Schematic of the tunneling junction device structure and atomic stacking of the oxide heterostructure. 
b) Expected effect of Zeeman splitting on the spin-dependent DOS for two cases: zero spin-orbit coupling ($b=0$) (left panel) and large spin-orbit coupling ($b=6$) (right panel). The dimensionless SOC parameter $b = \hbar/(3\tau_{so}\Delta_0)$ reflects the strength of the SOC relative to the gap energy scale. Dashed blue (dashed grey) and solid red (solid grey) curves represent the spin-up and spin-down DOS, respectively, while the solid black curve gives the total DOS from $\rho_{\uparrow} + \rho_{\downarrow}$ (shifted upwards by 1 for clarity). The spectra were calculated using Maki's theory (Eq. (\ref{Maki2})) at $T = 0$ K, lifetime broadening parameter $\zeta = 0.1$, and magnetic field $\mu_B H / \Delta_0 = 0.6$.
}
\end{figure}
%%%%%%%%%%%%
%%%%%%%%%%%%

The spin-orbit coupling strength can be quantitatively extracted from superconducting tunneling spectra of thin films in large parallel magnetic fields \cite{Meservey:1994, Fulde:1973, Meservey:1978}. 
In a conventional $s$-wave superconductor, a magnetic field acts in two ways on the conduction electrons:  by inducing cyclotron orbits and via the electron magnetic moment (spin). 
Both of these effects lead to the breaking of Cooper pairs once their energy scale competes with the condensation energy.  
For thin films in the 2D limit, the orbital depairing can be geometrically suppressed, leading to highly anisotropic upper-critical fields with large in-plane $H_{c2,\parallel}$. 
In the absence of spin-orbit coupling, spin is a good quantum number and $H_{c2,\parallel}$ is determined by the Pauli paramagnetic limit ($H_{P} =  \Delta_0 / \sqrt{2} \mu_B$, where $\Delta_0$ is the superconducting gap at $T = 0$ and $\mu_B$ is the Bohr magneton) \cite{Meservey:1994, Clogston:1962, Chandrasekhar:1962}. 
The application of an in-plane magnetic field splits the spin-up and spin-down superconducting quasiparticle density of states (DOS) through the Zeeman effect (Fig. 1 left panel) \cite{Meservey:1994}. 
Increasing the spin-orbit coupling leads to a mixing of the spin-up and spin-down states and lifts $H_{c2,\parallel}$ above the Pauli limit \cite{Meservey:1994, Wu:2006, Kim:2012}. 
If the spin-orbit scattering rate is very fast ($\hbar/\tau_{so} > \Delta_0$, where $\tau_{so}$ is the normal-state spin-orbit scattering time), then the superconducting DOS does not exhibit measurable Zeeman splitting (Fig. 1 right panel). 
Fitting the tunneling spectra using Maki's theory \cite{Maki:1964, Alexander:1985, Worledge:2000} enables a quantitative extraction of both the orbital depairing parameter ($\alpha_o$) and $\tau_{so}$ from the tunneling spectra. This approach, pioneered by Tedrow and Meservey, has been used extensively to explore depairing mechanisms of conventional elemental superconductors \cite{Meservey:1994, Meservey:1978, Fulde:1973, Worledge:2000, Alexander:1985}.

Here we examine spin-orbit coupling and orbital depairing in thin films of Nb-doped SrTiO$_3$ (NSTO) using tunneling spectroscopy. 
Recently, we have developed an approach for realizing high-quality tunneling junctions for bulk NSTO with $\mu$eV resolution of the superconducting gap \cite{Swartz:2018, Inoue:2015}. By carefully engineering the band alignments using polar tunneling barriers, the interfacial carrier density probed by tunneling corresponds to the nominal density of dopants.
We study the tunneling conductance ($di/dv$) of NSTO films in the 2D limit ($d < \xi_{GL}$, where $d$ is the film thickness and $\xi_{GL}$ is the Ginzburg-Landau coherence length). 
We find a single superconducting gap which closes at the superconducting transition temperature ($T_c$). 
We extract $H_{c2,\parallel}$ from the tunneling spectra and find that it greatly exceeds the Pauli limit.
Under in-plane applied fields, Zeeman splitting is not observed and an apparent single gap persists at all fields until closing completely near 1.6 T, indicating that the spin-orbit coupling scale ($\hbar/\tau_{so}$) is larger than $\Delta_0$.  
We analyze the data using Maki's theory \cite{Maki:1964, Worledge:2000, Alexander:1985} and examine the relative contributions from orbital depairing and spin-orbit scattering. 
Due to the heavy mixing of the spin states, Maki's theory provides an upper-bound for the spin-orbit scattering time of $\tau_{so} \leq 1.1$ ps and spin diffusion length $\lambda_s \leq 32$ nm. 

We fabricated tunneling junctions consisting of superconducting NSTO thin films of thickness $d$ = 18 nm, with a 2 unit cell (u.c.) epitaxial LaAlO$_3$ tunneling barrier, and Ag counter electrodes as described elsewhere \cite{Inoue:2015,Swartz:2018}.  
NSTO with 1 at.\% Nb-doping was homoepitaxially deposited on undoped SrTiO$_3$(001) by pulsed-laser deposition \cite{Kozuka:2010b}. 
Films grown by this technique exhibit full carrier activation and bulk-like electron mobility.  
The polar LaAlO$_3$ tunnel barrier plays a crucial role in enabling access to the electronic structure of NSTO in the 2D superconducting limit. 
The LaAlO$_3$ layer provides an interfacial electric dipole which shifts the band alignments between the Ag electrode and semiconducting SrTiO$_3$ by $\approx$$0.5$ eV/u.c. \cite{Yajima:2015, Tachikawa:2015, Hikita:2016}.
Aligning the Fermi-level between the two electrodes significantly reduces the Schottky barrier and eliminates the long depletion length which prohibits direct tunneling. 

First, we report the zero-field superconducting behavior of the sample. Figure 2a shows $di/dv$ measured at base temperature ($T$ = 20 mK) and $\mu_0 H$ = \mbox{0 T} exhibiting a single superconducting gap ($\Delta$). 
Although we observe high-energy coupling to longitudinal-optic phonon modes (not shown) as reported recently \cite{Swartz:2018}, we do not find other strong-coupling renormalizations (i.e. McMillan and Rowell \cite{McMillan:1965}) in the tunneling spectra. 
The superconducting gap is well fit by the Bardeen-Cooper-Schrieffer (BCS) equation for the density of states with $\Delta_0 = 47 \pm 1$ $\mu$eV. 
Due to the finite resolution of the measurement and thermal broadening, the minimum of the superconducting gap is finite. 
Here, the gap minimum is two-orders of magnitude smaller than the normal state conductance, demonstrating the dominance of elastic tunneling and the high quality of the junction, even in the 2D limit.  The superconducting gap closes near $T_c$ = 330 mK as measured by four-point resistivity (Fig. 2b). 
Importantly, we do not observe a pseudogap as was recently observed in LaAlO$_3$/SrTiO$_3$ \cite{Richter:2013}, indicating the pseudogap is specific to the LAO/STO interface and not a generic feature in the 2D limit. 

%%%%%%%%%%%%Figure 2
%%%%%%%%%%%%
\begin{figure}[t]
\includegraphics[width=85mm]{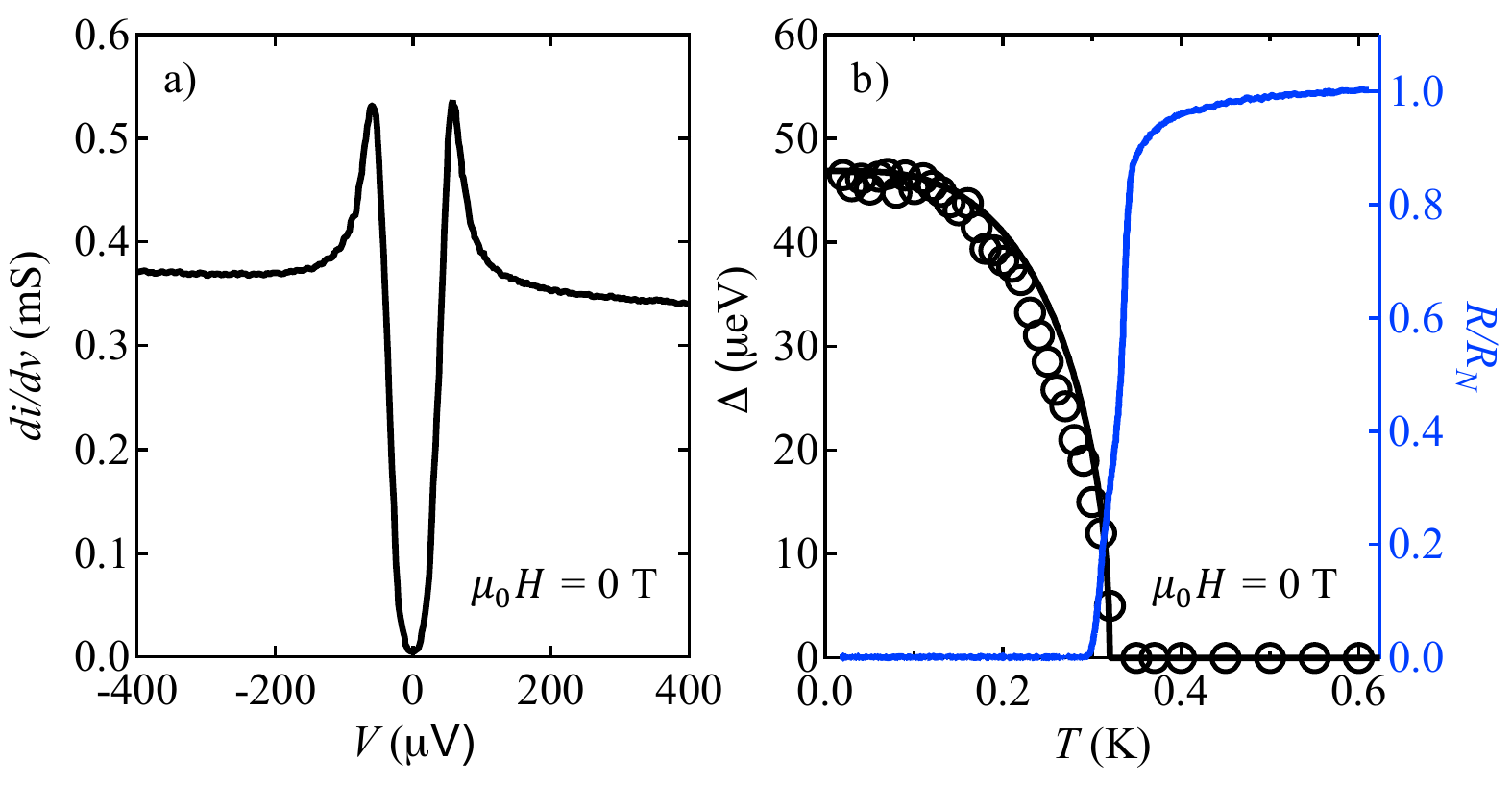}
\caption{\label{fig2} 
(Color online) Tunneling spectroscopy and resistivity in zero field. 
a) Tunneling conductance ($di/dv$) of 18 nm thick Nb-doped SrTiO$_3$ thin film measured at the base temperature of the dilution refrigerator. 
b) Superconducting gap amplitude ($\Delta$) (open circles, left axis) compared to the normalized resistance (solid blue (grey) line, right axis). The superconducting gap closes at $T = 315 \pm 5$ mK, which is very close to the resistive transition temperature $T_c$ = 330 mK defined as 50\% of the normal state resistivity at $T = 0.6$ K. 
}
\end{figure}
%%%%%%%%%%%%
%%%%%%%%%%%%

We now turn to the magnetic-field response of the superconducting gap. Figure 3a shows the superconducting gap at several characteristic values of applied magnetic field (left panel: $H_{\perp}$, right panel: $H_{\parallel}$). 
Figure 3b displays the zero-bias conductance (gap minimum) normalized to the normal-state zero-bias conductance for both field orientations. We find a large anisotropy between $H_{c2,\perp}$ and $H_{c2,\parallel}$ with a ratio $H_{c2,\perp}$ / $H_{c2,\parallel}$ = 0.052. 
We extract the Ginzburg-Landau superconducting coherence length $\xi_{GL} = \sqrt{\Phi_0 / (2 \pi H_{c2,\perp}})$ = 62 nm $>$ $d$, confirming the superconducting state is in the 2D regime.
SrTiO$_3$ is a type-II superconductor with large London penetration depth compared to $\xi_{GL}$ and $d$, and the quenching of superconductivity due to an out-of-plane field can be attributed to the formation of vortices. 
For fields applied in-plane, the large size of a vortex core is energetically unfavorable to form in the 2D limit and the orbital depairing is dramatically suppressed leading to enhanced $H_{c2,\parallel}$. 
We find that the superconducting gap exhibits large $H_{c2,\parallel}$ far in excess of the Pauli limit ($H_{P} =  \Delta_0 / \sqrt{2} \mu_B$ = 0.574 T), and in agreement with a study of upper-critical fields from resistivity measurements in $\delta$-doped SrTiO$_3$ quantum wells \cite{Kim:2012}.  Here, we can examine the spin-dependent response of the superconducting gap spectra to extract the relevant contributions to orbital and spin depairing mechanisms. 

%%%%%%%%%%%%Figure 3 
%%%%%%%%%%%%
\begin{figure}[t]
\includegraphics[width=85mm]{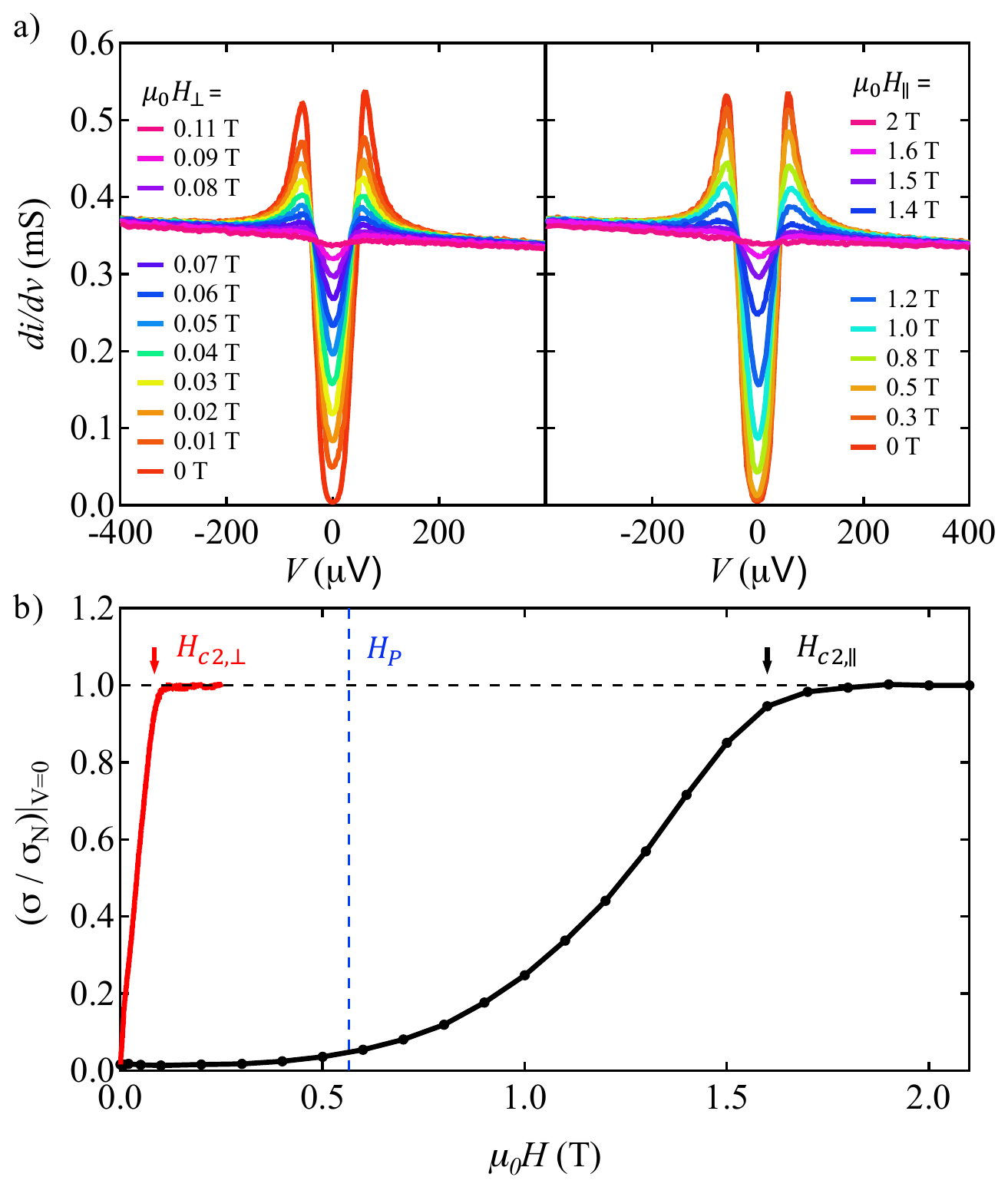}
\caption{\label{fig3} 
(Color online) Tunneling spectroscopy of the superconducting gap under applied magnetic field. 
a) Raw $di/dv$ data for several values of magnetic fields applied out-of plane ($\mu_0 H_{\perp}$, left panel) and in-plane ($\mu_0 H_{\parallel}$, right panel).  
b) Zero-bias conductivity ($\sigma = di/dv$) of the gap minimum normalized to the normal-state conductance ($\sigma_N$) for both field orientations. The out-of plane ($H_{c2,\perp}$) and in-plane ($H_{c2,\parallel}$) upper critical fields are indicated. The vertical dashed blue line indicates the Pauli paramagnetic limiting field ($H_P$).
}
\end{figure}
%%%%%%%%%%%%
%%%%%%%%%%%%

The superconducting DOS has been given by Maki's theory, which takes into account orbital depairing, Zeeman splitting of the spin states, and SOC \cite{Maki:1964, Fulde:1973}. The spin-dependent DOS is given by, 
\begin{equation}
\label{Maki1}
\rho_{\uparrow,\downarrow} = \frac{\rho_0}{2} \textrm{sgn}(E) \textrm{Re} \! \left( \frac{u_{\pm}}{\sqrt{u_{\pm}^2 - 1}}   \right),
\end{equation}
where $\rho_0$ is the normal-state DOS and $u_{\pm}$ are defined by,
\begin{equation}
\label{Maki2}
u_{\pm} = \frac{E\mp \mu_B H}{\Delta_0} + 
\zeta \frac{u_{\pm}}{\sqrt{1-u_{\pm}^2}} +
b \left(    \frac{u_{\mp}-u_{\pm}}{\sqrt{1-u_{\mp}^2}}                    \right),
\end{equation}
for which $E$ is the energy relative to the Fermi level ($E_F$), $b = \hbar/(3\tau_{so}\Delta_0$) is a dimensionless quantity representing the strength of the spin-orbit scattering relative to $\Delta_0$, and $\zeta$ represents spin-independent lifetime corrections.  
Maki's equation (Eq. (\ref{Maki2})) reduces to the BCS DOS in the limit of vanishing $\zeta$ and $b$. 
The quantity $\mu_B H$ represents the Zeeman splitting of the spin-dependent states and observation of this splitting in the experimental data depends on the strength of $b$ (see Fig. 1).
The parameter $\zeta = \alpha_i + \alpha_o H_{\parallel}^2$ includes field-independent broadening ($\alpha_i$) and $\alpha_o$ = $De^2d^2 / (6\hbar \Delta_0)$ is the standard orbital depairing for a thin film in a parallel magnetic field ($D$ is the diffusion coefficient) \cite{Maki:1964, Fulde:1973, Meservey:1994}. 
We follow the numerical approach of Worledge and Geballe in applying Eq. (\ref{Maki2}) to the tunneling data \cite{Maki:1964, Worledge:2000, Alexander:1985}. 

We now focus on the spectra shown in Fig. 3a (right panel) for in-plane applied fields. 
The magnetic fields explored here ($\mu_B H_{\parallel}/\Delta_0 < 2$) are large enough to observe Zeeman splitting in the weak spin-orbit limit ($b<1$) \cite{supplement}.
However, for all measured magnetic fields, the data does not exhibit a clear signature of Zeeman splitting indicating strong spin scattering relative to the superconducting gap (compare Fig. 1 with Fig. 3a right panel) and consistent with the violation of the Pauli-limit. 
While the spin-orbit parameter $b$ is field-independent, the effect of Zeeman splitting in combination with rapid spin mixing is to produce an \textit{effective} broadening of the total DOS ($\rho_{\uparrow} + \rho_{\downarrow}$, see Fig. 1) following an $H^2$ dependence \cite{Fulde:1973}. 
Therefore since both orbital depairing and the large SOC produce quasiparticle broadening under an applied field, it is a useful exercise to first consider a reduced version of Maki's theory which ignores the spin-degree of freedom in the problem, such that,  
\begin{equation}
\label{Maki_reduced}
u_{\pm} \rightarrow u = \frac{E}{\Delta_0} + 
\zeta' \frac{u}{\sqrt{1-u^2}},
\end{equation}
which in zero-field is equivalent to the Dynes formulation were the phenomenological Dynes quasiparticle broadening parameter is given by $\Gamma = \zeta' \Delta_0$ \cite{Dynes:1978}.  
We first fit the data of Fig. 3a right panel using Eq. (\ref{Maki_reduced}) where $\zeta'$ is the only free parameter. 
The results for $\zeta'$ are shown in Fig. 4a as a function of $H_{\parallel}^2$ and are well described by $\zeta' = \alpha_i + \eta H^2$ with $\alpha_i$ = 0.056 and $\eta$ = 0.4 T$^{-2}$. The small intrinsic quasiparticle broadening ($\alpha_i$) gives $\Gamma$ = 2 $\mu$eV and identical to our previous report in the bulk limit \cite{Swartz:2018}. 
The extracted $\eta$ value reflects the total contribution to field-induced broadening from both spin-orbit coupling and orbital depairing. 

To quantify the spin-orbit and orbital depairing contributions, we apply Maki's full theory (Eq. (\ref{Maki2})) to the set of tunneling data between 300 and 700 mT ($ 0.3 < \mu_B H / \Delta_0 < 0.86$) including the spin-dependent density of states, spin-orbit parameter $b$, and depairing parameter $\zeta = \alpha_i + \alpha_o H^2$.  
The only free parameters are $b$ and $\alpha_o$ which must both be singly valued at all fields. 
We find that the best fits are statistically equivalent for $b > 4$ (with varying $\alpha_0$) \cite{supplement}, indicating short spin-orbit scattering times $\tau_{so} < 1.1$ ps. 
In this regime ($b >4$), $\alpha_0$ and $b$ are correlated.
This can be understood as a competition between the spin-orbit induced effective broadening and orbital depairing. 
For instance, in the limit $b \rightarrow \infty$, the broadening from SOC vanishes and orbital depairing must asymptotically approach $\eta$ to account for the experimentally observed broadening. 
Fig. 4b shows a characteristic best fit for $\mu_0 H_{\parallel}$ = 0.5 T ($\mu_B H / \Delta_0$ = 0.61) with $b = 6$ and $\alpha_0 = 0.11$.
Additionally, an upper-bound on the spin diffusion length is given by $\lambda_s = \sqrt{\frac{3}{4} D_{tr} \tau_{so}} < 32$ nm \cite{Zutic:2004}, where $D_{tr} = v_F^2 \tau_{tr} / 3 \approx 0.0012$ m$^2$/s is the transport diffusion coefficient. 
Here we have estimated the Fermi velocity $v_F$ in a single-band approximation with effective mass $m^*$ = 1.24$m_0$ \cite{Kim:2011} and Fermi level $E_F$ = 61 meV \cite{Mazin:2011, Swartz:2018}.  
We have used the Drude scattering time $\tau_{tr} = m^* \mu_e / e$ where $\mu_e = 300$ cm$^2$/Vs is the experimentally measured electron mobility.

%%%%%%%%%%%%Figure 3 
%%%%%%%%%%%%
\begin{figure}[t]
\includegraphics[width=85mm]{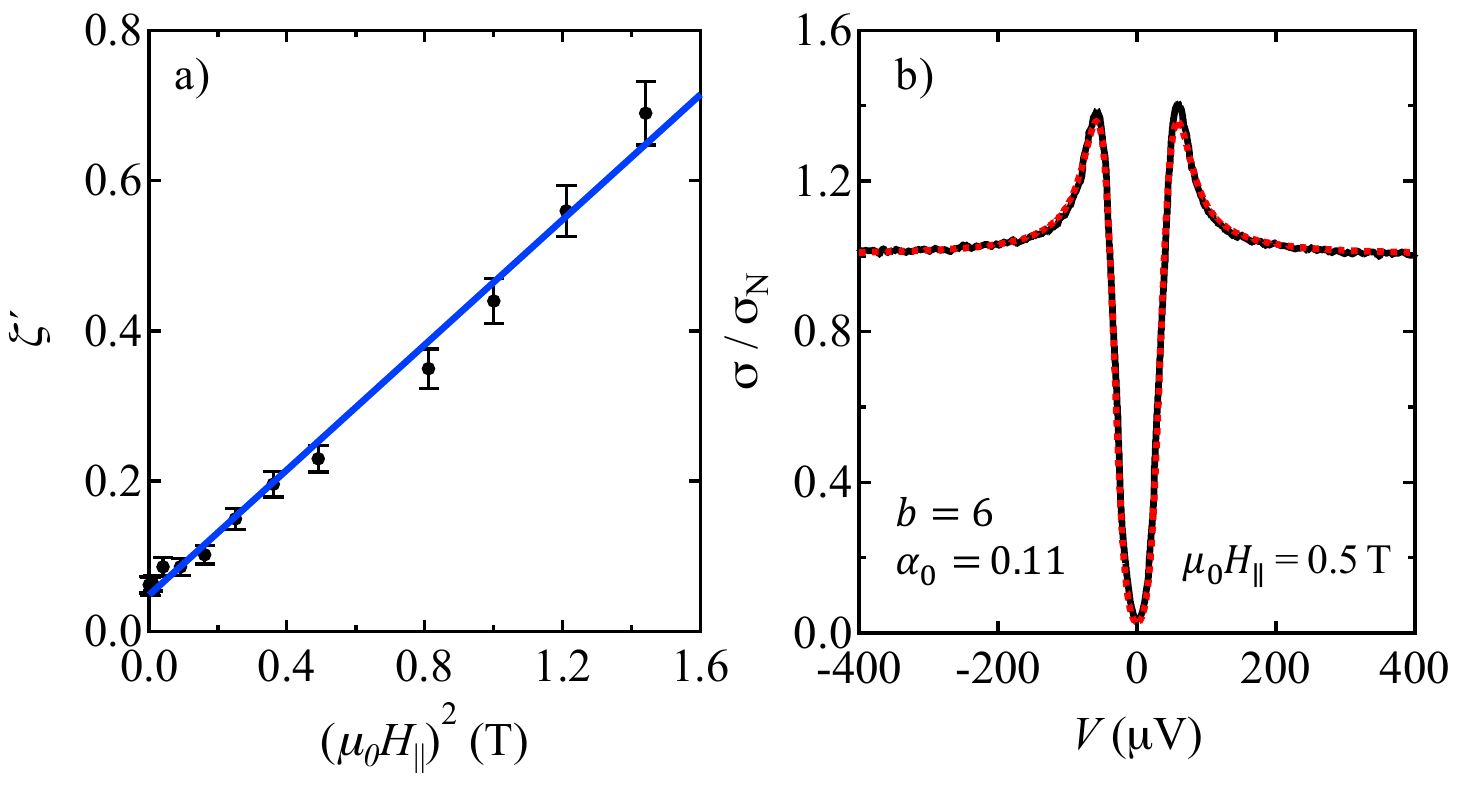}
\caption{\label{fig4} 
(Color online) Maki analysis of the superconducting gap spectra under in-plane magnetic fields. 
a) Total quasiparticle broadening $\zeta'$ (black dots) determined by fitting the tunneling data of Fig. 3a right panel using Eq. (\ref{Maki_reduced}). The total broadening exhibits a dependence on the square of the applied magnetic field and the solid line represents a fit to $\zeta' = \alpha_i + \eta H^2$.
b) Normalized $di/dv$ data (solid black line) measured at $\mu_0 H_{\parallel}$ = 0.5 T and theoretical fit (dashed red (grey) line) using Maki's full theory as expressed in Eq. (\ref{Maki2}) with $b$ = 6, $\alpha_i$ = 0.056, $\alpha_o$ = 0.11, and $\Delta_0$ = 47 $\mu$eV. 
}
\end{figure}
%%%%%%%%%%%%
%%%%%%%%%%%%

The contribution from orbital depairing in the tunneling data provides additional information on the superconducting phase. 
The best fits from Maki's theory in the range $b > 4$ correspond to 0.016 T$^{-2} < \alpha_o \leq \eta$, for which $\alpha_o$ increases commensurately with $b$. 
Thus, even though spin-orbit and orbital depairing cannot be quantified independently, there are clear experimental limits on $\alpha_o$.
We can compare the experimental $\alpha_o$ with the expected orbital contribution from normal-state transport parameters with $\alpha_o$ = $D_{tr}e^2d^2 / (6\hbar \Delta_0) \approx 2$ T$^{-2}$, which is far in excess of the measured total broadening of $\eta = 0.4$ T$^{-2}$. 
This apparent discrepancy can be resolved by considering the multi-band nature of bulk SrTiO$_3$ with three occupied $t_{2g}$ orbitals comprised of two light- and one heavy-electron bands \cite{Mazin:2011, Zhong:2013}. 
Normal-state transport coefficients are dominated by the highly mobile light electrons, but these carriers only make-up a fraction of the total DOS, whereas the lowest lying heavy band comprises the majority of the electrons at $E_F$ \cite{Mazin:2011, Zhong:2013, Kim:2011, Caviglia:2010}. 
In other words, the experimental data cannot be explained by solely considering highly mobile, light electrons in forming the superconducting phase.
We can transpose the orbital depairing extracted from the superconducting tunneling data to $D_{SC}$ representing the diffusion coefficient for electrons which contribute to pairing. We find $0.1\times10^{-4}$ m$^2$/s $< D_{SC} < 2.3\times 10^{-4}$ m$^2$/s, which agrees very well with a simplistic estimate of the diffusion constant for the heavy electron band with $m^* \approx 6 m_0$ \cite{Chang:2010} and momentum scattering time $\tau_{he} \approx 100$ fs, giving $D_{he} \approx 1 \times10^{-4}$ m$^2$/s. 
Therefore, the robustness of superconductivity at high magnetic fields is consistent with the established bulk band structure for which the heavy electron band dominates the total DOS and results in weak orbital depairing.  
We note that the importance of the heavy bands for superconductivity has been suggested in LaAlO$_3$/SrTiO$_3$ \cite{Joshua:2012, Nakamura:2013}

The spin-orbit scattering times observed here are comparable to the momentum scattering time ($\tau_{tr} / \tau_{so} \sim 0.1$) and significantly shorter than those suggested theoretically in a single band limit \cite{Sahin:2014}. Additionally, we can expect that Rashba and Dresselhaus fields are minimal in the current sample structure under investigation \cite{supplement}.
Therefore, the rapid spin mixing near the Fermi level can be understood in the context of the multiband electronic structure of bulk SrTiO$_3$ with hybridized orbital character arising from the tetragonal crystal field splitting and the intrinsic atomic spin-orbit interaction \cite{Mazin:2011}.
This picture is analogous to $p$-type Si where short spin relaxation times are characteristic despite the modest SOC \cite{Ando:2012, Jansen:2012b}.  
The spin-orbit scattering explored here reflects the electrons with the largest contribution to the density of states and the superconducting condensate, which in bulk SrTiO$_3$ is the heavy electron band.
This is in contrast to transport experiments exploring spin-orbit coupling in the normal state (i.e. weak (anti-)localization, Subnikov de Haas oscillations) which are most sensitive to the highly mobile subset of carriers \cite{Kim:2011, Caviglia:2010}.
Therefore, careful analysis of the sub-band structure and orbital character in confined SrTiO$_3$-based heterostructures (e.g. LaAlO$_3$/SrTiO$_3$) is critical to understanding the spin-orbit properties of the normal and superconducting phases. 
Regardless, it is interesting to note that the scattering times found here ($\tau_{so} \sim 1$ ps), are in the ballpark of the vast majority of experimental findings in LaAlO$_3$/SrTiO$_3$ \cite{Ben:2010, Caviglia:2010b}, suggesting that the spin-orbit scattering at the Fermi level arising from the intrinsic atomic spin-orbit interaction contributes at least on equal footing with Rashba effects. 

In conclusion, we have performed tunneling experiments on the dilute superconductor SrTiO$_3$ doped with 1 at.\% Nb in the 2D superconducting limit. 
These results were enabled by precisely designing the tunneling junction with epitaxial dipole tunnel barriers, which shift band alignments and facilitates high-resolution tunneling spectroscopy. 
The data indicates a single superconducting gap which closes at $T_c$. 
By geometrically suppressing the orbital depairing, we show that the large intrinsic SOC can be observed directly in the tunneling spectra by the violation of the Pauli-limit and the absence of Zeeman splitting. 
Surprisingly short spin-orbit scattering times of order 1 ps were obtained.
Examination of the orbital depairing parameter indicates that the heavy electron band, which is difficult to explore in transport experiments, plays an important role in the formation of the superconducting phase.

We thank M. E. Flatt\'{e} for useful discussions.
This work was supported by the Department of Energy, Office of Basic Energy Sciences, Division of Materials Sciences and Engineering, under Contract No. DE-AC02-76SF00515; and the Gordon and Betty Moore Foundation's EPiQS Initiative through Grant GBMF4415 (dilution fridge measurements). \\


\begin{thebibliography}{41}%
\makeatletter
\providecommand \@ifxundefined [1]{%
 \@ifx{#1\undefined}
}%
\providecommand \@ifnum [1]{%
 \ifnum #1\expandafter \@firstoftwo
 \else \expandafter \@secondoftwo
 \fi
}%
\providecommand \@ifx [1]{%
 \ifx #1\expandafter \@firstoftwo
 \else \expandafter \@secondoftwo
 \fi
}%
\providecommand \natexlab [1]{#1}%
\providecommand \enquote  [1]{``#1''}%
\providecommand \bibnamefont  [1]{#1}%
\providecommand \bibfnamefont [1]{#1}%
\providecommand \citenamefont [1]{#1}%
\providecommand \href@noop [0]{\@secondoftwo}%
\providecommand \href [0]{\begingroup \@sanitize@url \@href}%
\providecommand \@href[1]{\@@startlink{#1}\@@href}%
\providecommand \@@href[1]{\endgroup#1\@@endlink}%
\providecommand \@sanitize@url [0]{\catcode `\\12\catcode `\$12\catcode
  `\&12\catcode `\#12\catcode `\^12\catcode `\_12\catcode `\%12\relax}%
\providecommand \@@startlink[1]{}%
\providecommand \@@endlink[0]{}%
\providecommand \url  [0]{\begingroup\@sanitize@url \@url }%
\providecommand \@url [1]{\endgroup\@href {#1}{\urlprefix }}%
\providecommand \urlprefix  [0]{URL }%
\providecommand \Eprint [0]{\href }%
\providecommand \doibase [0]{http://dx.doi.org/}%
\providecommand \selectlanguage [0]{\@gobble}%
\providecommand \bibinfo  [0]{\@secondoftwo}%
\providecommand \bibfield  [0]{\@secondoftwo}%
\providecommand \translation [1]{[#1]}%
\providecommand \BibitemOpen [0]{}%
\providecommand \bibitemStop [0]{}%
\providecommand \bibitemNoStop [0]{.\EOS\space}%
\providecommand \EOS [0]{\spacefactor3000\relax}%
\providecommand \BibitemShut  [1]{\csname bibitem#1\endcsname}%
\let\auto@bib@innerbib\@empty
%</preamble>
\bibitem [{\citenamefont {Winkler}()}]{Winkler:2003}%
  \BibitemOpen
  \bibfield  {author} {\bibinfo {author} {\bibfnamefont {R.}~\bibnamefont
  {Winkler}},\ }\href@noop {} {\emph {\bibinfo {title} {Spin-Orbit Coupling
  Effects in Two-Dimensional Electron and Hole Systems}}}\BibitemShut {NoStop}%
\bibitem [{\citenamefont {van~der Marel}\ \emph {et~al.}(2011)\citenamefont
  {van~der Marel}, \citenamefont {van Mechelen},\ and\ \citenamefont
  {Mazin}}]{Mazin:2011}%
  \BibitemOpen
  \bibfield  {author} {\bibinfo {author} {\bibfnamefont {D.}~\bibnamefont
  {van~der Marel}}, \bibinfo {author} {\bibfnamefont {J.~L.~M.}\ \bibnamefont
  {van Mechelen}}, \ and\ \bibinfo {author} {\bibfnamefont {I.~I.}\
  \bibnamefont {Mazin}},\ }\href@noop {} {\bibfield  {journal} {\bibinfo
  {journal} {Phys. Rev. B}\ }\textbf {\bibinfo {volume} {84}},\ \bibinfo
  {pages} {205111} (\bibinfo {year} {2011})}\BibitemShut {NoStop}%
\bibitem [{\citenamefont {Kim}\ \emph {et~al.}(2011)\citenamefont {Kim},
  \citenamefont {Bell}, \citenamefont {Kozuka}, \citenamefont {Kurita},
  \citenamefont {Hikita},\ and\ \citenamefont {Hwang}}]{Kim:2011}%
  \BibitemOpen
  \bibfield  {author} {\bibinfo {author} {\bibfnamefont {M.}~\bibnamefont
  {Kim}}, \bibinfo {author} {\bibfnamefont {C.}~\bibnamefont {Bell}}, \bibinfo
  {author} {\bibfnamefont {Y.}~\bibnamefont {Kozuka}}, \bibinfo {author}
  {\bibfnamefont {M.}~\bibnamefont {Kurita}}, \bibinfo {author} {\bibfnamefont
  {Y.}~\bibnamefont {Hikita}}, \ and\ \bibinfo {author} {\bibfnamefont {H.~Y.}\
  \bibnamefont {Hwang}},\ }\href@noop {} {\bibfield  {journal} {\bibinfo
  {journal} {Phys. Rev. Lett.}\ }\textbf {\bibinfo {volume} {107}},\ \bibinfo
  {pages} {106801} (\bibinfo {year} {2011})}\BibitemShut {NoStop}%
\bibitem [{\citenamefont {Lin}\ \emph {et~al.}(2013)\citenamefont {Lin},
  \citenamefont {Zhu}, \citenamefont {Fauqu\'e},\ and\ \citenamefont
  {Behnia}}]{Lin:2013}%
  \BibitemOpen
  \bibfield  {author} {\bibinfo {author} {\bibfnamefont {X.}~\bibnamefont
  {Lin}}, \bibinfo {author} {\bibfnamefont {Z.}~\bibnamefont {Zhu}}, \bibinfo
  {author} {\bibfnamefont {B.}~\bibnamefont {Fauqu\'e}}, \ and\ \bibinfo
  {author} {\bibfnamefont {K.}~\bibnamefont {Behnia}},\ }\href@noop {}
  {\bibfield  {journal} {\bibinfo  {journal} {Phys. Rev. X}\ }\textbf {\bibinfo
  {volume} {3}},\ \bibinfo {pages} {021002} (\bibinfo {year}
  {2013})}\BibitemShut {NoStop}%
\bibitem [{\citenamefont {Qing-Yan}\ \emph {et~al.}(2012)\citenamefont
  {Qing-Yan}, \citenamefont {Zhi}, \citenamefont {Wen-Hao}, \citenamefont
  {Zuo-Cheng}, \citenamefont {Jin-Song}, \citenamefont {Wei}, \citenamefont
  {Hao}, \citenamefont {Yun-Bo}, \citenamefont {Peng}, \citenamefont {Kai}
  \emph {et~al.}}]{Qing:2012}%
  \BibitemOpen
  \bibfield  {author} {\bibinfo {author} {\bibfnamefont {W.}~\bibnamefont
  {Qing-Yan}}, \bibinfo {author} {\bibfnamefont {L.}~\bibnamefont {Zhi}},
  \bibinfo {author} {\bibfnamefont {Z.}~\bibnamefont {Wen-Hao}}, \bibinfo
  {author} {\bibfnamefont {Z.}~\bibnamefont {Zuo-Cheng}}, \bibinfo {author}
  {\bibfnamefont {Z.}~\bibnamefont {Jin-Song}}, \bibinfo {author}
  {\bibfnamefont {L.}~\bibnamefont {Wei}}, \bibinfo {author} {\bibfnamefont
  {D.}~\bibnamefont {Hao}}, \bibinfo {author} {\bibfnamefont {O.}~\bibnamefont
  {Yun-Bo}}, \bibinfo {author} {\bibfnamefont {D.}~\bibnamefont {Peng}},
  \bibinfo {author} {\bibfnamefont {C.}~\bibnamefont {Kai}},  \emph {et~al.},\
  }\href@noop {} {\bibfield  {journal} {\bibinfo  {journal} {Chin. Phys.
  Lett.}\ }\textbf {\bibinfo {volume} {29}},\ \bibinfo {pages} {037402}
  (\bibinfo {year} {2012})}\BibitemShut {NoStop}%
\bibitem [{\citenamefont {Caviglia}\ \emph {et~al.}(2008)\citenamefont
  {Caviglia}, \citenamefont {Gariglio}, \citenamefont {Reyren}, \citenamefont
  {Jaccard}, \citenamefont {Schneider}, \citenamefont {Gabay}, \citenamefont
  {Thiel}, \citenamefont {Hammerl}, \citenamefont {Mannhart},\ and\
  \citenamefont {Triscone}}]{Caviglia:2008}%
  \BibitemOpen
  \bibfield  {author} {\bibinfo {author} {\bibfnamefont {A.~D.}\ \bibnamefont
  {Caviglia}}, \bibinfo {author} {\bibfnamefont {S.}~\bibnamefont {Gariglio}},
  \bibinfo {author} {\bibfnamefont {N.}~\bibnamefont {Reyren}}, \bibinfo
  {author} {\bibfnamefont {D.}~\bibnamefont {Jaccard}}, \bibinfo {author}
  {\bibfnamefont {T.}~\bibnamefont {Schneider}}, \bibinfo {author}
  {\bibfnamefont {M.}~\bibnamefont {Gabay}}, \bibinfo {author} {\bibfnamefont
  {S.}~\bibnamefont {Thiel}}, \bibinfo {author} {\bibfnamefont
  {G.}~\bibnamefont {Hammerl}}, \bibinfo {author} {\bibfnamefont
  {J.}~\bibnamefont {Mannhart}}, \ and\ \bibinfo {author} {\bibfnamefont
  {J.-M.}\ \bibnamefont {Triscone}},\ }\href@noop {} {\bibfield  {journal}
  {\bibinfo  {journal} {Nature}\ }\textbf {\bibinfo {volume} {456}},\ \bibinfo
  {pages} {624} (\bibinfo {year} {2008})}\BibitemShut {NoStop}%
\bibitem [{\citenamefont {Ben~Shalom}\ \emph {et~al.}(2010)\citenamefont
  {Ben~Shalom}, \citenamefont {Sachs}, \citenamefont {Rakhmilevitch},
  \citenamefont {Palevski},\ and\ \citenamefont {Dagan}}]{Ben:2010}%
  \BibitemOpen
  \bibfield  {author} {\bibinfo {author} {\bibfnamefont {M.}~\bibnamefont
  {Ben~Shalom}}, \bibinfo {author} {\bibfnamefont {M.}~\bibnamefont {Sachs}},
  \bibinfo {author} {\bibfnamefont {D.}~\bibnamefont {Rakhmilevitch}}, \bibinfo
  {author} {\bibfnamefont {A.}~\bibnamefont {Palevski}}, \ and\ \bibinfo
  {author} {\bibfnamefont {Y.}~\bibnamefont {Dagan}},\ }\href@noop {}
  {\bibfield  {journal} {\bibinfo  {journal} {Phys. Rev. Lett.}\ }\textbf
  {\bibinfo {volume} {104}},\ \bibinfo {pages} {126802} (\bibinfo {year}
  {2010})}\BibitemShut {NoStop}%
\bibitem [{\citenamefont {Caviglia}\ \emph
  {et~al.}(2010{\natexlab{a}})\citenamefont {Caviglia}, \citenamefont {Gabay},
  \citenamefont {Gariglio}, \citenamefont {Reyren}, \citenamefont
  {Cancellieri},\ and\ \citenamefont {Triscone}}]{Caviglia:2010b}%
  \BibitemOpen
  \bibfield  {author} {\bibinfo {author} {\bibfnamefont {A.~D.}\ \bibnamefont
  {Caviglia}}, \bibinfo {author} {\bibfnamefont {M.}~\bibnamefont {Gabay}},
  \bibinfo {author} {\bibfnamefont {S.}~\bibnamefont {Gariglio}}, \bibinfo
  {author} {\bibfnamefont {N.}~\bibnamefont {Reyren}}, \bibinfo {author}
  {\bibfnamefont {C.}~\bibnamefont {Cancellieri}}, \ and\ \bibinfo {author}
  {\bibfnamefont {J.-M.}\ \bibnamefont {Triscone}},\ }\href@noop {} {\bibfield
  {journal} {\bibinfo  {journal} {Phys. Rev. Lett.}\ }\textbf {\bibinfo
  {volume} {104}},\ \bibinfo {pages} {126803} (\bibinfo {year}
  {2010}{\natexlab{a}})}\BibitemShut {NoStop}%
\bibitem [{\citenamefont {Nakamura}\ \emph {et~al.}(2012)\citenamefont
  {Nakamura}, \citenamefont {Koga},\ and\ \citenamefont
  {Kimura}}]{Nakamura:2012}%
  \BibitemOpen
  \bibfield  {author} {\bibinfo {author} {\bibfnamefont {H.}~\bibnamefont
  {Nakamura}}, \bibinfo {author} {\bibfnamefont {T.}~\bibnamefont {Koga}}, \
  and\ \bibinfo {author} {\bibfnamefont {T.}~\bibnamefont {Kimura}},\
  }\href@noop {} {\bibfield  {journal} {\bibinfo  {journal} {Phys. Rev. Lett.}\
  }\textbf {\bibinfo {volume} {108}},\ \bibinfo {pages} {206601} (\bibinfo
  {year} {2012})}\BibitemShut {NoStop}%
\bibitem [{\citenamefont {Zhong}\ \emph {et~al.}(2013)\citenamefont {Zhong},
  \citenamefont {T{\'o}th},\ and\ \citenamefont {Held}}]{Zhong:2013}%
  \BibitemOpen
  \bibfield  {author} {\bibinfo {author} {\bibfnamefont {Z.}~\bibnamefont
  {Zhong}}, \bibinfo {author} {\bibfnamefont {A.}~\bibnamefont {T{\'o}th}}, \
  and\ \bibinfo {author} {\bibfnamefont {K.}~\bibnamefont {Held}},\ }\href@noop
  {} {\bibfield  {journal} {\bibinfo  {journal} {Phys. Rev. B}\ }\textbf
  {\bibinfo {volume} {87}},\ \bibinfo {pages} {161102} (\bibinfo {year}
  {2013})}\BibitemShut {NoStop}%
\bibitem [{\citenamefont {Khalsa}\ \emph {et~al.}(2012)\citenamefont {Khalsa},
  \citenamefont {Lee},\ and\ \citenamefont {MacDonald}}]{Khalsa:2013}%
  \BibitemOpen
  \bibfield  {author} {\bibinfo {author} {\bibfnamefont {G.}~\bibnamefont
  {Khalsa}}, \bibinfo {author} {\bibfnamefont {B.}~\bibnamefont {Lee}}, \ and\
  \bibinfo {author} {\bibfnamefont {A.~H.}\ \bibnamefont {MacDonald}},\
  }\href@noop {} {\bibfield  {journal} {\bibinfo  {journal} {Phys. Rev. B}\
  }\textbf {\bibinfo {volume} {86}},\ \bibinfo {pages} {125121} (\bibinfo
  {year} {2012})}\BibitemShut {NoStop}%
\bibitem [{\citenamefont {King}\ \emph {et~al.}(2014)\citenamefont {King},
  \citenamefont {Walker}, \citenamefont {Tamai}, \citenamefont {De~La~Torre},
  \citenamefont {Eknapakul}, \citenamefont {Buaphet}, \citenamefont {Mo},
  \citenamefont {Meevasana}, \citenamefont {Bahramy},\ and\ \citenamefont
  {Baumberger}}]{King:2014}%
  \BibitemOpen
  \bibfield  {author} {\bibinfo {author} {\bibfnamefont {P.}~\bibnamefont
  {King}}, \bibinfo {author} {\bibfnamefont {S.~M.}\ \bibnamefont {Walker}},
  \bibinfo {author} {\bibfnamefont {A.}~\bibnamefont {Tamai}}, \bibinfo
  {author} {\bibfnamefont {A.}~\bibnamefont {De~La~Torre}}, \bibinfo {author}
  {\bibfnamefont {T.}~\bibnamefont {Eknapakul}}, \bibinfo {author}
  {\bibfnamefont {P.}~\bibnamefont {Buaphet}}, \bibinfo {author} {\bibfnamefont
  {S.}~\bibnamefont {Mo}}, \bibinfo {author} {\bibfnamefont {W.}~\bibnamefont
  {Meevasana}}, \bibinfo {author} {\bibfnamefont {M.}~\bibnamefont {Bahramy}},
  \ and\ \bibinfo {author} {\bibfnamefont {F.}~\bibnamefont {Baumberger}},\
  }\href@noop {} {\bibfield  {journal} {\bibinfo  {journal} {Nature Commun.}\
  }\textbf {\bibinfo {volume} {5}},\ \bibinfo {pages} {3414} (\bibinfo {year}
  {2014})}\BibitemShut {NoStop}%
\bibitem [{\citenamefont {Rout}\ \emph {et~al.}(2017)\citenamefont {Rout},
  \citenamefont {Maniv},\ and\ \citenamefont {Dagan}}]{Rout:2017}%
  \BibitemOpen
  \bibfield  {author} {\bibinfo {author} {\bibfnamefont {P.~K.}\ \bibnamefont
  {Rout}}, \bibinfo {author} {\bibfnamefont {E.}~\bibnamefont {Maniv}}, \ and\
  \bibinfo {author} {\bibfnamefont {Y.}~\bibnamefont {Dagan}},\ }\href@noop {}
  {\bibfield  {journal} {\bibinfo  {journal} {Phys. Rev. Lett.}\ }\textbf
  {\bibinfo {volume} {119}},\ \bibinfo {pages} {237002} (\bibinfo {year}
  {2017})}\BibitemShut {NoStop}%
\bibitem [{\citenamefont {Meservey}\ and\ \citenamefont
  {Tedrow}(1994)}]{Meservey:1994}%
  \BibitemOpen
  \bibfield  {author} {\bibinfo {author} {\bibfnamefont {R.}~\bibnamefont
  {Meservey}}\ and\ \bibinfo {author} {\bibfnamefont {P.}~\bibnamefont
  {Tedrow}},\ }\href@noop {} {\bibfield  {journal} {\bibinfo  {journal} {Phys.
  Rep.}\ }\textbf {\bibinfo {volume} {238}},\ \bibinfo {pages} {173} (\bibinfo
  {year} {1994})}\BibitemShut {NoStop}%
\bibitem [{\citenamefont {Fulde}(1973)}]{Fulde:1973}%
  \BibitemOpen
  \bibfield  {author} {\bibinfo {author} {\bibfnamefont {P.}~\bibnamefont
  {Fulde}},\ }\href@noop {} {\bibfield  {journal} {\bibinfo  {journal} {Adv.
  Phys.}\ }\textbf {\bibinfo {volume} {22}},\ \bibinfo {pages} {667} (\bibinfo
  {year} {1973})}\BibitemShut {NoStop}%
\bibitem [{\citenamefont {Meservey}\ \emph {et~al.}(1978)\citenamefont
  {Meservey}, \citenamefont {Tedrow},\ and\ \citenamefont
  {Bruno}}]{Meservey:1978}%
  \BibitemOpen
  \bibfield  {author} {\bibinfo {author} {\bibfnamefont {R.}~\bibnamefont
  {Meservey}}, \bibinfo {author} {\bibfnamefont {P.}~\bibnamefont {Tedrow}}, \
  and\ \bibinfo {author} {\bibfnamefont {R.~C.}\ \bibnamefont {Bruno}},\
  }\href@noop {} {\bibfield  {journal} {\bibinfo  {journal} {Phys. Rev. B}\
  }\textbf {\bibinfo {volume} {17}},\ \bibinfo {pages} {2915} (\bibinfo {year}
  {1978})}\BibitemShut {NoStop}%
\bibitem [{\citenamefont {Clogston}(1962)}]{Clogston:1962}%
  \BibitemOpen
  \bibfield  {author} {\bibinfo {author} {\bibfnamefont {A.~M.}\ \bibnamefont
  {Clogston}},\ }\href@noop {} {\bibfield  {journal} {\bibinfo  {journal}
  {Phys. Rev. Lett.}\ }\textbf {\bibinfo {volume} {9}},\ \bibinfo {pages} {266}
  (\bibinfo {year} {1962})}\BibitemShut {NoStop}%
\bibitem [{\citenamefont {Chandrasekhar}(1962)}]{Chandrasekhar:1962}%
  \BibitemOpen
  \bibfield  {author} {\bibinfo {author} {\bibfnamefont {B.}~\bibnamefont
  {Chandrasekhar}},\ }\href@noop {} {\bibfield  {journal} {\bibinfo  {journal}
  {Appl. Phys. Lett.}\ }\textbf {\bibinfo {volume} {1}},\ \bibinfo {pages} {7}
  (\bibinfo {year} {1962})}\BibitemShut {NoStop}%
\bibitem [{\citenamefont {Wu}\ \emph {et~al.}(2006)\citenamefont {Wu},
  \citenamefont {Adams}, \citenamefont {Yang},\ and\ \citenamefont
  {McCarley}}]{Wu:2006}%
  \BibitemOpen
  \bibfield  {author} {\bibinfo {author} {\bibfnamefont {X.}~\bibnamefont
  {Wu}}, \bibinfo {author} {\bibfnamefont {P.}~\bibnamefont {Adams}}, \bibinfo
  {author} {\bibfnamefont {Y.}~\bibnamefont {Yang}}, \ and\ \bibinfo {author}
  {\bibfnamefont {R.}~\bibnamefont {McCarley}},\ }\href@noop {} {\bibfield
  {journal} {\bibinfo  {journal} {Phys. Rev. Lett.}\ }\textbf {\bibinfo
  {volume} {96}},\ \bibinfo {pages} {127002} (\bibinfo {year}
  {2006})}\BibitemShut {NoStop}%
\bibitem [{\citenamefont {Kim}\ \emph {et~al.}(2012)\citenamefont {Kim},
  \citenamefont {Kozuka}, \citenamefont {Bell}, \citenamefont {Hikita},\ and\
  \citenamefont {Hwang}}]{Kim:2012}%
  \BibitemOpen
  \bibfield  {author} {\bibinfo {author} {\bibfnamefont {M.}~\bibnamefont
  {Kim}}, \bibinfo {author} {\bibfnamefont {Y.}~\bibnamefont {Kozuka}},
  \bibinfo {author} {\bibfnamefont {C.}~\bibnamefont {Bell}}, \bibinfo {author}
  {\bibfnamefont {Y.}~\bibnamefont {Hikita}}, \ and\ \bibinfo {author}
  {\bibfnamefont {H.}~\bibnamefont {Hwang}},\ }\href@noop {} {\bibfield
  {journal} {\bibinfo  {journal} {Phys. Rev. B}\ }\textbf {\bibinfo {volume}
  {86}},\ \bibinfo {pages} {085121} (\bibinfo {year} {2012})}\BibitemShut
  {NoStop}%
\bibitem [{\citenamefont {Maki}(1964)}]{Maki:1964}%
  \BibitemOpen
  \bibfield  {author} {\bibinfo {author} {\bibfnamefont {K.}~\bibnamefont
  {Maki}},\ }\href@noop {} {\bibfield  {journal} {\bibinfo  {journal} {Pog.
  Theor. Phys.}\ }\textbf {\bibinfo {volume} {32}},\ \bibinfo {pages} {29}
  (\bibinfo {year} {1964})}\BibitemShut {NoStop}%
\bibitem [{\citenamefont {Alexander}\ \emph {et~al.}(1985)\citenamefont
  {Alexander}, \citenamefont {Orlando}, \citenamefont {Rainer},\ and\
  \citenamefont {Tedrow}}]{Alexander:1985}%
  \BibitemOpen
  \bibfield  {author} {\bibinfo {author} {\bibfnamefont {J.}~\bibnamefont
  {Alexander}}, \bibinfo {author} {\bibfnamefont {T.}~\bibnamefont {Orlando}},
  \bibinfo {author} {\bibfnamefont {D.}~\bibnamefont {Rainer}}, \ and\ \bibinfo
  {author} {\bibfnamefont {P.}~\bibnamefont {Tedrow}},\ }\href@noop {}
  {\bibfield  {journal} {\bibinfo  {journal} {Phys. Rev. B}\ }\textbf {\bibinfo
  {volume} {31}},\ \bibinfo {pages} {5811} (\bibinfo {year}
  {1985})}\BibitemShut {NoStop}%
\bibitem [{\citenamefont {Worledge}\ and\ \citenamefont
  {Geballe}(2000)}]{Worledge:2000}%
  \BibitemOpen
  \bibfield  {author} {\bibinfo {author} {\bibfnamefont {D.}~\bibnamefont
  {Worledge}}\ and\ \bibinfo {author} {\bibfnamefont {T.}~\bibnamefont
  {Geballe}},\ }\href@noop {} {\bibfield  {journal} {\bibinfo  {journal} {Phys.
  Rev. B}\ }\textbf {\bibinfo {volume} {62}},\ \bibinfo {pages} {447} (\bibinfo
  {year} {2000})}\BibitemShut {NoStop}%
\bibitem [{\citenamefont {Swartz}\ \emph {et~al.}(2018)\citenamefont {Swartz},
  \citenamefont {Inoue}, \citenamefont {Merz}, \citenamefont {Hikita},
  \citenamefont {Raghu}, \citenamefont {Devereaux}, \citenamefont {Johnston},\
  and\ \citenamefont {Hwang}}]{Swartz:2018}%
  \BibitemOpen
  \bibfield  {author} {\bibinfo {author} {\bibfnamefont {A.~G.}\ \bibnamefont
  {Swartz}}, \bibinfo {author} {\bibfnamefont {H.}~\bibnamefont {Inoue}},
  \bibinfo {author} {\bibfnamefont {T.~A.}\ \bibnamefont {Merz}}, \bibinfo
  {author} {\bibfnamefont {Y.}~\bibnamefont {Hikita}}, \bibinfo {author}
  {\bibfnamefont {S.}~\bibnamefont {Raghu}}, \bibinfo {author} {\bibfnamefont
  {T.~P.}\ \bibnamefont {Devereaux}}, \bibinfo {author} {\bibfnamefont
  {S.}~\bibnamefont {Johnston}}, \ and\ \bibinfo {author} {\bibfnamefont
  {H.~Y.}\ \bibnamefont {Hwang}},\ }\href@noop {} {\bibfield  {journal}
  {\bibinfo  {journal} {Proc. Natl. Acad. Sci.}\ }\textbf {\bibinfo {volume}
  {115}},\ \bibinfo {pages} {1475} (\bibinfo {year} {2018})}\BibitemShut
  {NoStop}%
\bibitem [{\citenamefont {Inoue}\ \emph {et~al.}(2015)\citenamefont {Inoue},
  \citenamefont {Swartz}, \citenamefont {Harmon}, \citenamefont {Tachikawa},
  \citenamefont {Hikita}, \citenamefont {Flatt\'e},\ and\ \citenamefont
  {Hwang}}]{Inoue:2015}%
  \BibitemOpen
  \bibfield  {author} {\bibinfo {author} {\bibfnamefont {H.}~\bibnamefont
  {Inoue}}, \bibinfo {author} {\bibfnamefont {A.~G.}\ \bibnamefont {Swartz}},
  \bibinfo {author} {\bibfnamefont {N.~J.}\ \bibnamefont {Harmon}}, \bibinfo
  {author} {\bibfnamefont {T.}~\bibnamefont {Tachikawa}}, \bibinfo {author}
  {\bibfnamefont {Y.}~\bibnamefont {Hikita}}, \bibinfo {author} {\bibfnamefont
  {M.~E.}\ \bibnamefont {Flatt\'e}}, \ and\ \bibinfo {author} {\bibfnamefont
  {H.~Y.}\ \bibnamefont {Hwang}},\ }\href@noop {} {\bibfield  {journal}
  {\bibinfo  {journal} {Phys. Rev. X}\ }\textbf {\bibinfo {volume} {5}},\
  \bibinfo {pages} {041023} (\bibinfo {year} {2015})}\BibitemShut {NoStop}%
\bibitem [{\citenamefont {Kozuka}\ \emph {et~al.}(2010)\citenamefont {Kozuka},
  \citenamefont {Hikita}, \citenamefont {Bell},\ and\ \citenamefont
  {Hwang}}]{Kozuka:2010b}%
  \BibitemOpen
  \bibfield  {author} {\bibinfo {author} {\bibfnamefont {Y.}~\bibnamefont
  {Kozuka}}, \bibinfo {author} {\bibfnamefont {Y.}~\bibnamefont {Hikita}},
  \bibinfo {author} {\bibfnamefont {C.}~\bibnamefont {Bell}}, \ and\ \bibinfo
  {author} {\bibfnamefont {H.~Y.}\ \bibnamefont {Hwang}},\ }\href@noop {}
  {\bibfield  {journal} {\bibinfo  {journal} {Appl. Phys. Lett.}\ }\textbf
  {\bibinfo {volume} {97}},\ \bibinfo {eid} {012107} (\bibinfo {year}
  {2010})}\BibitemShut {NoStop}%
\bibitem [{\citenamefont {Yajima}\ \emph {et~al.}(2015)\citenamefont {Yajima},
  \citenamefont {Minohara}, \citenamefont {Bell}, \citenamefont {Kumigashira},
  \citenamefont {Oshima}, \citenamefont {Hwang},\ and\ \citenamefont
  {Hikita}}]{Yajima:2015}%
  \BibitemOpen
  \bibfield  {author} {\bibinfo {author} {\bibfnamefont {T.}~\bibnamefont
  {Yajima}}, \bibinfo {author} {\bibfnamefont {M.}~\bibnamefont {Minohara}},
  \bibinfo {author} {\bibfnamefont {C.}~\bibnamefont {Bell}}, \bibinfo {author}
  {\bibfnamefont {H.}~\bibnamefont {Kumigashira}}, \bibinfo {author}
  {\bibfnamefont {M.}~\bibnamefont {Oshima}}, \bibinfo {author} {\bibfnamefont
  {H.~Y.}\ \bibnamefont {Hwang}}, \ and\ \bibinfo {author} {\bibfnamefont
  {Y.}~\bibnamefont {Hikita}},\ }\href@noop {} {\bibfield  {journal} {\bibinfo
  {journal} {Nano Lett.}\ }\textbf {\bibinfo {volume} {15}},\ \bibinfo {pages}
  {1622} (\bibinfo {year} {2015})}\BibitemShut {NoStop}%
\bibitem [{\citenamefont {Tachikawa}\ \emph {et~al.}(2015)\citenamefont
  {Tachikawa}, \citenamefont {Minohara}, \citenamefont {Hikita}, \citenamefont
  {Bell},\ and\ \citenamefont {Hwang}}]{Tachikawa:2015}%
  \BibitemOpen
  \bibfield  {author} {\bibinfo {author} {\bibfnamefont {T.}~\bibnamefont
  {Tachikawa}}, \bibinfo {author} {\bibfnamefont {M.}~\bibnamefont {Minohara}},
  \bibinfo {author} {\bibfnamefont {Y.}~\bibnamefont {Hikita}}, \bibinfo
  {author} {\bibfnamefont {C.}~\bibnamefont {Bell}}, \ and\ \bibinfo {author}
  {\bibfnamefont {H.~Y.}\ \bibnamefont {Hwang}},\ }\href@noop {} {\bibfield
  {journal} {\bibinfo  {journal} {Adv. Mater.}\ }\textbf {\bibinfo {volume}
  {27}},\ \bibinfo {pages} {7458} (\bibinfo {year} {2015})}\BibitemShut
  {NoStop}%
\bibitem [{\citenamefont {Hikita}\ \emph {et~al.}(2016)\citenamefont {Hikita},
  \citenamefont {Nishio}, \citenamefont {Seitz}, \citenamefont {Chakthranont},
  \citenamefont {Tachikawa}, \citenamefont {Jaramillo},\ and\ \citenamefont
  {Hwang}}]{Hikita:2016}%
  \BibitemOpen
  \bibfield  {author} {\bibinfo {author} {\bibfnamefont {Y.}~\bibnamefont
  {Hikita}}, \bibinfo {author} {\bibfnamefont {K.}~\bibnamefont {Nishio}},
  \bibinfo {author} {\bibfnamefont {L.~C.}\ \bibnamefont {Seitz}}, \bibinfo
  {author} {\bibfnamefont {P.}~\bibnamefont {Chakthranont}}, \bibinfo {author}
  {\bibfnamefont {T.}~\bibnamefont {Tachikawa}}, \bibinfo {author}
  {\bibfnamefont {T.~F.}\ \bibnamefont {Jaramillo}}, \ and\ \bibinfo {author}
  {\bibfnamefont {H.~Y.}\ \bibnamefont {Hwang}},\ }\href@noop {} {\bibfield
  {journal} {\bibinfo  {journal} {Adv. Energy Mater.}\ }\textbf {\bibinfo
  {volume} {6}},\ \bibinfo {pages} {1502154} (\bibinfo {year}
  {2016})}\BibitemShut {NoStop}%
\bibitem [{\citenamefont {McMillan}\ and\ \citenamefont
  {Rowell}(1965)}]{McMillan:1965}%
  \BibitemOpen
  \bibfield  {author} {\bibinfo {author} {\bibfnamefont {W.~L.}\ \bibnamefont
  {McMillan}}\ and\ \bibinfo {author} {\bibfnamefont {J.~M.}\ \bibnamefont
  {Rowell}},\ }\href@noop {} {\bibfield  {journal} {\bibinfo  {journal} {Phys.
  Rev. Lett.}\ }\textbf {\bibinfo {volume} {14}},\ \bibinfo {pages} {108}
  (\bibinfo {year} {1965})}\BibitemShut {NoStop}%
\bibitem [{\citenamefont {Richter}\ \emph {et~al.}(2013)\citenamefont
  {Richter}, \citenamefont {Boschker}, \citenamefont {Dietsche}, \citenamefont
  {Fillis-Tsirakis}, \citenamefont {Jany}, \citenamefont {Loder}, \citenamefont
  {Kourkoutis}, \citenamefont {Muller}, \citenamefont {Kirtley}, \citenamefont
  {Schneider} \emph {et~al.}}]{Richter:2013}%
  \BibitemOpen
  \bibfield  {author} {\bibinfo {author} {\bibfnamefont {C.}~\bibnamefont
  {Richter}}, \bibinfo {author} {\bibfnamefont {H.}~\bibnamefont {Boschker}},
  \bibinfo {author} {\bibfnamefont {W.}~\bibnamefont {Dietsche}}, \bibinfo
  {author} {\bibfnamefont {E.}~\bibnamefont {Fillis-Tsirakis}}, \bibinfo
  {author} {\bibfnamefont {R.}~\bibnamefont {Jany}}, \bibinfo {author}
  {\bibfnamefont {F.}~\bibnamefont {Loder}}, \bibinfo {author} {\bibfnamefont
  {L.}~\bibnamefont {Kourkoutis}}, \bibinfo {author} {\bibfnamefont
  {D.}~\bibnamefont {Muller}}, \bibinfo {author} {\bibfnamefont
  {J.}~\bibnamefont {Kirtley}}, \bibinfo {author} {\bibfnamefont
  {C.}~\bibnamefont {Schneider}},  \emph {et~al.},\ }\href@noop {} {\bibfield
  {journal} {\bibinfo  {journal} {Nature}\ }\textbf {\bibinfo {volume} {502}},\
  \bibinfo {pages} {528} (\bibinfo {year} {2013})}\BibitemShut {NoStop}%
\bibitem [{sup()}]{supplement}%
  \BibitemOpen
  \href@noop {} {\bibinfo  {journal} {See Supplementary Information}\
  }\BibitemShut {NoStop}%
\bibitem [{\citenamefont {Dynes}\ \emph {et~al.}(1978)\citenamefont {Dynes},
  \citenamefont {Narayanamurti},\ and\ \citenamefont {Garno}}]{Dynes:1978}%
  \BibitemOpen
\bibfield  {journal} {  }\bibfield  {author} {\bibinfo {author} {\bibfnamefont
  {R.}~\bibnamefont {Dynes}}, \bibinfo {author} {\bibfnamefont
  {V.}~\bibnamefont {Narayanamurti}}, \ and\ \bibinfo {author} {\bibfnamefont
  {J.~P.}\ \bibnamefont {Garno}},\ }\href@noop {} {\bibfield  {journal}
  {\bibinfo  {journal} {Phys. Rev. Lett.}\ }\textbf {\bibinfo {volume} {41}},\
  \bibinfo {pages} {1509} (\bibinfo {year} {1978})}\BibitemShut {NoStop}%
\bibitem [{\citenamefont {{\v{Z}}uti{\'c}}\ \emph {et~al.}(2004)\citenamefont
  {{\v{Z}}uti{\'c}}, \citenamefont {Fabian},\ and\ \citenamefont
  {Sarma}}]{Zutic:2004}%
  \BibitemOpen
  \bibfield  {author} {\bibinfo {author} {\bibfnamefont {I.}~\bibnamefont
  {{\v{Z}}uti{\'c}}}, \bibinfo {author} {\bibfnamefont {J.}~\bibnamefont
  {Fabian}}, \ and\ \bibinfo {author} {\bibfnamefont {S.~D.}\ \bibnamefont
  {Sarma}},\ }\href@noop {} {\bibfield  {journal} {\bibinfo  {journal} {Rev.
  Mod. Phys.}\ }\textbf {\bibinfo {volume} {76}},\ \bibinfo {pages} {323}
  (\bibinfo {year} {2004})}\BibitemShut {NoStop}%
\bibitem [{\citenamefont {Caviglia}\ \emph
  {et~al.}(2010{\natexlab{b}})\citenamefont {Caviglia}, \citenamefont
  {Gariglio}, \citenamefont {Cancellieri}, \citenamefont {Sac{\'e}p{\'e}},
  \citenamefont {Fete}, \citenamefont {Reyren}, \citenamefont {Gabay},
  \citenamefont {Morpurgo},\ and\ \citenamefont {Triscone}}]{Caviglia:2010}%
  \BibitemOpen
  \bibfield  {author} {\bibinfo {author} {\bibfnamefont {A.~D.}\ \bibnamefont
  {Caviglia}}, \bibinfo {author} {\bibfnamefont {S.}~\bibnamefont {Gariglio}},
  \bibinfo {author} {\bibfnamefont {C.}~\bibnamefont {Cancellieri}}, \bibinfo
  {author} {\bibfnamefont {B.}~\bibnamefont {Sac{\'e}p{\'e}}}, \bibinfo
  {author} {\bibfnamefont {A.}~\bibnamefont {Fete}}, \bibinfo {author}
  {\bibfnamefont {N.}~\bibnamefont {Reyren}}, \bibinfo {author} {\bibfnamefont
  {M.}~\bibnamefont {Gabay}}, \bibinfo {author} {\bibfnamefont {A.~F.}\
  \bibnamefont {Morpurgo}}, \ and\ \bibinfo {author} {\bibfnamefont {J.-M.}\
  \bibnamefont {Triscone}},\ }\href@noop {} {\bibfield  {journal} {\bibinfo
  {journal} {Phys. Rev. Lett.}\ }\textbf {\bibinfo {volume} {105}},\ \bibinfo
  {pages} {236802} (\bibinfo {year} {2010}{\natexlab{b}})}\BibitemShut
  {NoStop}%
\bibitem [{\citenamefont {Chang}\ \emph {et~al.}(2010)\citenamefont {Chang},
  \citenamefont {Bostwick}, \citenamefont {Kim}, \citenamefont {Horn},\ and\
  \citenamefont {Rotenberg}}]{Chang:2010}%
  \BibitemOpen
  \bibfield  {author} {\bibinfo {author} {\bibfnamefont {Y.~J.}\ \bibnamefont
  {Chang}}, \bibinfo {author} {\bibfnamefont {A.}~\bibnamefont {Bostwick}},
  \bibinfo {author} {\bibfnamefont {Y.~S.}\ \bibnamefont {Kim}}, \bibinfo
  {author} {\bibfnamefont {K.}~\bibnamefont {Horn}}, \ and\ \bibinfo {author}
  {\bibfnamefont {E.}~\bibnamefont {Rotenberg}},\ }\href@noop {} {\bibfield
  {journal} {\bibinfo  {journal} {Phys. Rev. B}\ }\textbf {\bibinfo {volume}
  {81}},\ \bibinfo {pages} {235109} (\bibinfo {year} {2010})}\BibitemShut
  {NoStop}%
\bibitem [{\citenamefont {Joshua}\ \emph {et~al.}(2012)\citenamefont {Joshua},
  \citenamefont {Pecker}, \citenamefont {Ruhman}, \citenamefont {Altman},\ and\
  \citenamefont {Ilani}}]{Joshua:2012}%
  \BibitemOpen
  \bibfield  {author} {\bibinfo {author} {\bibfnamefont {A.}~\bibnamefont
  {Joshua}}, \bibinfo {author} {\bibfnamefont {S.}~\bibnamefont {Pecker}},
  \bibinfo {author} {\bibfnamefont {J.}~\bibnamefont {Ruhman}}, \bibinfo
  {author} {\bibfnamefont {E.}~\bibnamefont {Altman}}, \ and\ \bibinfo {author}
  {\bibfnamefont {S.}~\bibnamefont {Ilani}},\ }\href@noop {} {\bibfield
  {journal} {\bibinfo  {journal} {Nature Commun.}\ }\textbf {\bibinfo {volume}
  {3}},\ \bibinfo {pages} {1129} (\bibinfo {year} {2012})}\BibitemShut
  {NoStop}%
\bibitem [{\citenamefont {Nakamura}\ and\ \citenamefont
  {Yanase}(2013)}]{Nakamura:2013}%
  \BibitemOpen
  \bibfield  {author} {\bibinfo {author} {\bibfnamefont {Y.}~\bibnamefont
  {Nakamura}}\ and\ \bibinfo {author} {\bibfnamefont {Y.}~\bibnamefont
  {Yanase}},\ }\href@noop {} {\bibfield  {journal} {\bibinfo  {journal} {J.
  Phys. Soc. Jpn.}\ }\textbf {\bibinfo {volume} {82}},\ \bibinfo {pages}
  {083705} (\bibinfo {year} {2013})}\BibitemShut {NoStop}%
\bibitem [{\citenamefont {{\c{S}}ahin}\ \emph {et~al.}(2014)\citenamefont
  {{\c{S}}ahin}, \citenamefont {Vignale},\ and\ \citenamefont
  {Flatt{\'e}}}]{Sahin:2014}%
  \BibitemOpen
  \bibfield  {author} {\bibinfo {author} {\bibfnamefont {C.}~\bibnamefont
  {{\c{S}}ahin}}, \bibinfo {author} {\bibfnamefont {G.}~\bibnamefont
  {Vignale}}, \ and\ \bibinfo {author} {\bibfnamefont {M.~E.}\ \bibnamefont
  {Flatt{\'e}}},\ }\href@noop {} {\bibfield  {journal} {\bibinfo  {journal}
  {Phys. Rev. B}\ }\textbf {\bibinfo {volume} {89}},\ \bibinfo {pages} {155402}
  (\bibinfo {year} {2014})}\BibitemShut {NoStop}%
\bibitem [{\citenamefont {Ando}\ and\ \citenamefont
  {Saitoh}(2012)}]{Ando:2012}%
  \BibitemOpen
  \bibfield  {author} {\bibinfo {author} {\bibfnamefont {K.}~\bibnamefont
  {Ando}}\ and\ \bibinfo {author} {\bibfnamefont {E.}~\bibnamefont {Saitoh}},\
  }\href@noop {} {\bibfield  {journal} {\bibinfo  {journal} {Nature Commun.}\
  }\textbf {\bibinfo {volume} {3}},\ \bibinfo {pages} {629} (\bibinfo {year}
  {2012})}\BibitemShut {NoStop}%
\bibitem [{\citenamefont {Jansen}\ \emph {et~al.}(2012)\citenamefont {Jansen},
  \citenamefont {Dash}, \citenamefont {Sharma},\ and\ \citenamefont
  {Min}}]{Jansen:2012b}%
  \BibitemOpen
  \bibfield  {author} {\bibinfo {author} {\bibfnamefont {R.}~\bibnamefont
  {Jansen}}, \bibinfo {author} {\bibfnamefont {S.~P.}\ \bibnamefont {Dash}},
  \bibinfo {author} {\bibfnamefont {S.}~\bibnamefont {Sharma}}, \ and\ \bibinfo
  {author} {\bibfnamefont {B.~C.}\ \bibnamefont {Min}},\ }\href@noop {}
  {\bibfield  {journal} {\bibinfo  {journal} {Semicond. Sci. Technol.}\
  }\textbf {\bibinfo {volume} {27}},\ \bibinfo {pages} {083001} (\bibinfo
  {year} {2012})}\BibitemShut {NoStop}%
\end{thebibliography}
\end{document}